\documentclass[12pt]{article}

\usepackage{amsmath,amssymb}
\usepackage{graphicx}
\usepackage{dcolumn}
\usepackage{bm}
\usepackage[numbers,super,comma,sort&compress]{natbib}

\begin{document}

\makeatletter
  \renewcommand\@biblabel[1]{#1.}
  \makeatother

\bibliographystyle{apsrev}

\renewcommand{\baselinestretch}{1.0}
\normalsize

{\sffamily
\noindent {\bfseries \Large Scaling of the Strain Hardening Modulus of Glassy Polymers with the Flow Stress}\\

\noindent {Mark O. Robbins}\\
\noindent {Department of Physics and Astronomy, Johns Hopkins University, Baltimore, MD 21218}\\
\noindent {Robert S. Hoy}\\
\noindent {Materials Research Laboratory, University of California, Santa Barbara, CA 93106}\\
}

\vspace{-5mm}
\noindent {\em Dated: April 20, 2009}\\

\begin{center}
\begin{minipage}{5.5in}
{\bf ABSTRACT:}  
In a recent letter, Govaert et al. examined the relationship between
strain hardening modulus $G_r$ and flow stress $\sigma_{flow}$
for five different glassy polymers.
In each case, results for $G_r$ at different strain rates or different temperatures
were linearly related to the flow stress.
They suggested that this linear relation was inconsistent with
simulations.
Data from previous publications and new results are presented to show that
simulations also yield a linear relation between modulus and flow stress.
Possible explanations for the change in the ratio of modulus to flow stress with temperature and strain rate are discussed.


\end{minipage}
\end{center}

\section*{\sffamily \large INTRODUCTION}
\label{sec:Intro}
The response of polymers to large strains plays a key role in determining their failure mode and mechanical performance.
Strain hardening, an increase in stress with increasing strain, makes it more difficult to deform regions that have already yielded.
This spreads deformation to new regions, preventing the localization of strain that would lead to brittle fracture.

Many experimental measurements of strain hardening have been interpreted in terms of an entropic network model \cite{arruda93b} based on rubber elasticity theories.\cite{treloar75}
The idea is that the entanglements between chains act like chemical crosslinks.
The increase in stress is attributed to the decrease in entropy as the chain segments between these effective crosslinks are stretched.
While this theory describes the functional form of stress-strain curves,
the magnitude of the rate of stress increase, or hardening modulus $G_r$,
is hard to understand from this model.\cite{vanMelick03,kramer05}
Assuming that the entanglement density is comparable to that in the melt,
the hardening modulus should be comparable to the melt plateau modulus $G_{melt}$ near $T_g$, and decrease with decreasing temperature.
Instead $G_r$ is typically two orders of magnitude larger than $G_{melt}$ and {\it increases} with decreasing $T$.

In a series of recent papers,\cite{hoy06,hoy07,hoy08} we have used molecular simulations to probe the origins of strain hardening.
The microscopic behavior is not consistent with the entropic network model.
Instead, strain hardening is directly related to the rate of plastic rearrangements needed to maintain chain connectivity.
The scale of the hardening modulus $G_r$ is thus set by the flow stress $\sigma_{\rm flow}$ rather than entropic stresses.
In both these simulations \cite{hoy06,hoy07,hoy08} and experiments, \cite{vanMelick03,dupaix05,wendlandt05} the values of $G_r$ and $\sigma_{\rm flow}$ are of the same order of magnitude and both increase with decreasing temperature.
Our simulations used a coarse-grained bead-spring model,\cite{kremer90}
but qualitatively similar behavior is observed
with more realistic potentials.\cite{lyulin05}
A recent microscopic theory of glassy polymers also leads to similar
scaling of plastic flow and hardening stresses.\cite{chen09}

Motivated by these studies, Govaert et al.\cite{govaert08} examined
the hardening modulus and flow stress of five different glassy polymers
over a range of rates and temperatures.
They found a very interesting linear relationship between $G_r$ and
$\sigma_{flow}$ 
\begin{equation}
G_r = C_0 + C_1 \sigma_{flow}
\label{eq:c0c1}
\end{equation}
for all the polymers studied.
Unfortunately they took our statement that the yield stress sets the scale of the hardening modulus to mean that the ratio of hardening modulus to yield stress is strictly constant, and thus argued that the observation of 
a nonzero $C_0$ seemed inconsistent with our results.
This is not the case, but their work has inspired us to re-examine the relation
between $G_r$ and $\sigma_{flow}$.

In this paper we compare previously published simulation results \cite{hoy06,hoy07,hoy08} and new data to the experimental data shown by Govaert et al..\cite{govaert08}
All simulations are consistent with Eq.\ \ref{eq:c0c1}, and $C_0$ is not in general zero.
The implications of Eq.\ \ref{eq:c0c1} as the temperature $T$ approaches the
glass transition temperature $T_g$ are discussed, as well as possible origins
for $C_0$ due to variations in thermal activation with strain.

\section*{\sffamily \large SIMULATION METHODS}
\label{sec:Methods}

The simulations follow the methodology described in our previous
papers.\cite{hoy06,hoy07,hoy08}
A generic bead-spring model \cite{kremer90} that describes the coarse-grained behavior
of polymers is used.
Each polymer contains $N=350$ spherical beads that interact with a
truncated and shifted Lennard-Jones (LJ) potential $U_{LJ} = 4u_0\left[ (a/r)^{12} - (a/r)^{6} - (a/r_c)^{12} + (a/r_c)^6 \right]$.
The binding energy $u_0$ and 
molecular diameter $a$
are used to define our units.
The unit of time $\tau = \sqrt{ma^2/u_0}$, where $m$ is the bead mass.

Adjacent beads along the chain are coupled with the finitely extensible nonlinear elastic (FENE) potential, $U_{FENE}(r) = -kR_0^2 ln(1 - (r/R_0)^2)$, which prevents chain crossing and scission.
The standard \cite{kremer90} values $k = 30u_0/a^2$ and $R_0 = 1.5a$ are employed, giving an equilibrium bond length $l_0 \simeq 0.96a$.
The entanglement density is varied by adding a bending
potential $U_{bend}$ to change the chain stiffness.
This potential, $U_{bend} (\theta)=k_{bend}(1-\cos \theta)$, 
where $\theta$ is the angle between consecutive covalent bond vectors.
Increasing $k_{bend}$ from 0 to $2.0 u_{0}$ changes the number of monomers per entanglement length $N_e$ from about 70 to 20.\cite{everaers04,sukumaran05}

Cubic samples with $N_{ch}=200$ chains and periodic boundary conditions
are equilibrated using the double-bridging-MD hybrid (DBH) algorithm.\cite{auhl03}
A uniaxial compression is applied along the $z$ axis while maintaining
zero stress along the transverse $(x,y)$ directions.\cite{yang97}
The stretch $\lambda=L_z/L_{z}^{0}$, where $L_z$ is the period
along the $z$ direction and $L_{z}^{0}$ is the initial value.
A constant true strain rate $\dot{\epsilon} = \dot{\lambda}/\lambda$ is
applied.
The stress $\sigma$ along the compressive axis is plotted against the Green-Lagrange strain $g(\lambda) =\lambda^2-1/\lambda$,
since the slope of this curve corresponds to the hardening modulus
in entropic models and experimental analysis: $G_r \equiv \partial \sigma/\partial g$.

The strain rates $\dot{\epsilon}$ used here range from
$|\dot{\epsilon}| = 10^{-6} /\tau$ to $10^{-3}/\tau$.
Over this range we find no qualitative change in behavior, just a roughly logarithmic shift in stress with rate that is also seen in many experiments\cite{gsell79} and
is consistent with activated models\cite{ree55} for flow stress.
Since mappings of the bead-spring model to real systems give $\tau$ in the
picosecond to nanosecond range (e.\ g.\  66ps for PE\cite{kremer90}),
the slowest rates used here overlap with the highest rates accessible in experiments ($10^4$ to $10^5 s^{-1}$).\cite{mulliken06,sarva07}

In both simulations and experiments, the initial yield behavior is sensitive
to the age and preparation of the glassy system.\cite{arruda93b,klompen05}
Shear rejuvenates the system  and the behavior for $|g| > 0.5$ is
fairly independent of past history.
Experimental data typically show a large initial yield stress followed
by strain softening.
There is then a minimum stress before strain hardening sets in.
Govaert et al. associated this minimum stress with the flow stress.\cite{govaert08}
The value of $|g|$ at the minimum varies between about 0.5 and 1 with polymer and temperature, and the stress is fairly constant over this range of $|g|$.
In our simulations, the glass has not generally been aged long enough to produce
significant strain softening.
There is an initial elastic increase in stress followed by a nearly
constant plateau. 
We identified the stress at a point near the end of this plateau $|g|=0.5$
with the flow stress.
This is close to the minimum in the stress for simulations that do show strain softening.

Since the experiments and simulations deal with compression, the stress along the compressive axis is negative.
Following Govaert et al. we will take $\sigma_{flow}$ as a positive number
equal to the magnitude
of the stress.
As in Ref. \cite{hoy06}, 
the value of $G_r$ was obtained from a linear fit to $\sigma$ as a
function of $g$ from $g=-0.5$ to -3.
 
\section*{\sffamily \large RESULTS AND DISCUSSION}
\label{sec:Results}

\begin{figure}
\includegraphics[width=3.375in]{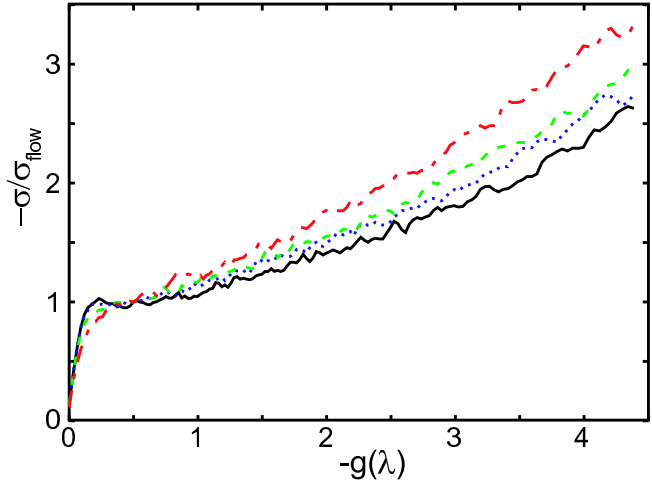}
\caption{(Color online)
Compressive stress $-\sigma$ normalized by $\sigma_{flow}$
as a function of $-g(\lambda)$
for systems with $k_{bend}=0.75u_0$ at $k_B T/\epsilon = 0.01$ (solid), 0.1 (dotted), 0.2 (dashed) and 0.3 (dash-dotted).
The strain rate $\dot{\epsilon}=-3.16 \cdot 10^{-4} \tau^{-1}$.
(Rescaled data from Fig. 3 of Ref. 5.)
}
\label{fig:flows}
\end{figure}

Figure \ref{fig:flows} shows plots of $\sigma/\sigma_{flow}$ for several temperatures and a fixed $\dot{\epsilon} \tau=-3.16\cdot 10^{-4}$.
Note that these normalized curves show an increase in slope
with increasing $T$,
implying that the ratio $G_r/\sigma_{flow}$
increases with increasing temperature.  
Using a value of $|g(\lambda)|$ other than 0.5 to define the flow stress would not change this trend.
Govaert et al. found a similar increase for poly(ethylene terphthalate)-glycol (PETG) and a smaller increase for polystyrene (PS).\cite{govaert08}
The trend with temperature had the opposite sign for polycarbonate (PC) and
poly(methyl methacrylate) (PMMA).

\begin{figure}
\includegraphics[width=3.375in]{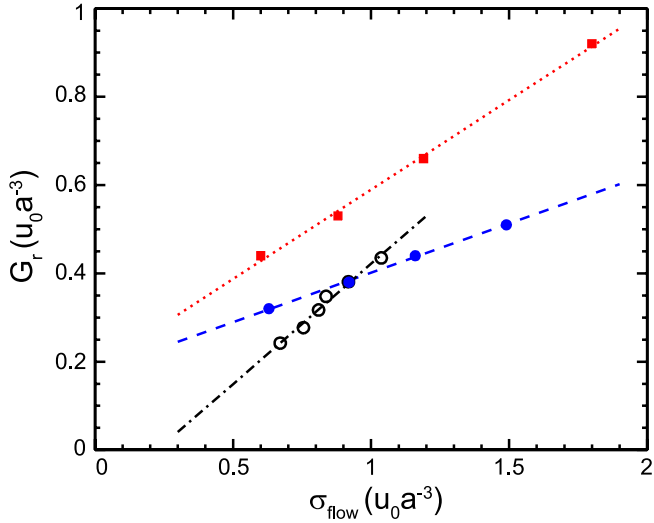}
\caption{(Color online)
Hardening modulus as a function of flow stress
for systems with $k_{bend}=0.75 u_0$ (circles) and $1.5 u_0$ (squares).
Filled symbols show results for fixed strain rate and
four values of the temperature, $k_B T/u_0 = 0.01$, 0.1, 0.2 and 0.3.
Open symbols show results for $k_B T/u_0 =0.2$ and six values of strain rate
$|\dot{\epsilon}| \tau =10^{-6},$ $10^{-5}$, $3.16\cdot 10^{-5}$, $10^{-4}$, $3.16 \cdot 10^{-4}$, and $10^{-3}$.
Both $G_r$ and $\sigma_{\rm flow}$ decrease with increasing temperature
and decreasing strain rate.
The straight lines are linear fits like those obtained in Ref. \cite{govaert08}.}
\label{fig:Grvssf}
\end{figure}

Given the observed change in $G_r/\sigma_{\rm flow}$, Govaert et al. examined the functional relation between the two quantities in more detail
and discovered that results for all polymers could be fit
to Eq. \ref{eq:c0c1}.
Figure \ref{fig:Grvssf} shows that our simulation results follow
the same linear relation.
Results for the temperature dependence at two values of chain stiffness are shown by solid symbols.
The two have very different slopes, but similar positive offsets.
We found that decreasing the strain rate by a factor of 30 only changed the
offset by about 40\%.
As expected from entropic models, and our picture of hardening through plastic
rearrangements, the slope is always steeper for more entangled polymers (bigger $k_{bend}$).
Note that the unit of stress \cite{rottler02}
 $u_0/a^3$ is of order 50MPa so
that the simulation results for $G_r$ and $\sigma_{flow}$ are quite
comparable to the experimental values (5 to 40MPa and 15 to 120MPa, respectively).

The $k_{bend}=0.75u_0$ data at fixed shear rate shown in Fig. \ref{fig:Grvssf}
(circles) was included in Table 2 of Ref. \cite{hoy06}.
This table also quoted values of $G_r/\sigma_{flow}$, which
systematically increased by 50\% with increasing $T$.
Given this, it is surprising that Govaert \textit{et.\ al.} concluded that our simulations gave a constant $G_r/\sigma_{flow}$.
Note that the ratio between $G_r$ and the entropic prediction changes by a
factor of 50 for the same data.
This is one reason for
our conclusion that the scale of $G_r$ is set by $\sigma_{flow}$
rather than the entropic stress.

The linear relation between $G_r$ and $\sigma_{\rm flow}$ leads to interesting behavior as $T$ increases to the glass transition temperature $T_g$ where $\sigma_{flow} \rightarrow 0$.
In Fig. \ref{fig:Grvssf}, the results for varying $T$ have $C_0 > 0$, implying that hardening would still be observed in the absence of a flow stress.
While a hardening modulus of order the melt plateau modulus would 
be expected in this limit, the values of $C_0$ in our simulations and
for PETG and PS are substantially larger than $G_{melt}$. 

One possible explanation of the large $C_0$ was mentioned in
the discussion of changes in $G_r/\sigma_{flow}$ in Ref. \cite{hoy06}.
Strain hardening is correlated with an increase in the size of regions that
must be plastically deformed to shear the system while retaining chain connectivity.
This is evidenced by an increase in the magnitude of non-affine deformations\cite{hoy06}
and in the number of Lennard Jones
bonds broken\cite{hoy07,hoy08} as $|g|$ increases.
The flow stress goes to zero when $T$ is high enough to activate local segment-scale rearrangements at the given strain rate.
This temperature will not in general be high enough to activate the larger,
correlated rearrangements needed to maintain chain connectivity at higher
strains,
leading to a nonzero $G_r$ as $\sigma_{flow}$ goes to zero.

Govaert et al. also found cases where fits to the temperature dependent
hardening gave negative $C_0$.
This implies that the hardening modulus vanishes and then becomes negative as $T$ approaches $T_g$.
This behavior seems counterintuitive, but the only case where the fit clearly indicates a negative $C_0$ is PC, and the data for it show a systematic curvature towards the origin.
It would be interesting to extend the experimental data for all these systems
toward $T_g$.
One possibility is that the strain softening contribution to $\sigma$ is
still important at the minimum in the stress that was used to define
experimental values of $\sigma_{flow}$.
Indeed, the minimum occurs where the strain softening and straing hardening
terms balance.
The strain softening term in the experimental data
varies in magnitude and extent with temperature,
as does the value of $|g|$ at the stress minimum.

The rate dependence of $G_R$ is also of interest.
In Ref. \cite{hoy06} we found that plots of $\sigma/\sigma_{flow}$ for different strain rates and fixed temperature collapsed fairly well onto a single curve,
implying a small $C_0$.
Govaert et al. found fairly good collapses in
similar plots for PS and PMMA, but a clear change
in normalized stress curves for PETG, PC and poly(phenylene ether) (PPE).
All but PETG had a negative $C_0$.
Inspired by the experiments, we have extended our results to strain rates that are thirty times lower than in Ref. \cite{hoy06}.
As shown by the open symbols in Fig. \ref{fig:Grvssf},
results for $G_r$ vs. $\sigma_{flow}$ fall on a straight line with a negative
offset.
Ref. \cite{hoy06} only included the four highest rates (and thus highest
$\sigma_{flow}$).
While the ratio of $G_r/\sigma_{flow}$ changes relatively little for these
four points, even these data are consistent with a negative $C_0$.

Rate dependence is often discussed in terms of Eyring-like models \cite{ree55, wendlandt05}
where the stress rises with rate as $(k_B T/V^*) \ln{\dot{\epsilon}}$.
Here $V^*$ is called the activation volume and represents the derivative of
the activation energy barrier with respect to stress.
Both experiments \cite{gsell79,wendlandt05} and simulations \cite{rottler03,hoy08} show a logarithmic dependence of stress on rate, with a strain dependent prefactor that increases (implying $V^*$ decreases) during strain hardening \cite{foot2}.
This would be consistent with the complex rearrangements at large strains being harder to activate with a simple compressive stress.
The more rapid rate dependence at high strains leads to a drop
in strain hardening with decreasing rate, and thus is consistent with
negative values of $C_0$.
The only system considered by Govaert et al. that did not have a negative
$C_0$
was PETG, which showed little change in either $G_R$ or $\sigma_{flow}$.
Note that a negative $C_0$ is not problematic for the rate dependence since
it is not generally possible to bring either $G_R$ or $\sigma_{flow}$ 
to zero because of the weak logarithmic dependence on rate.

Govaert et al. state that the variation of $G_r/\sigma_{flow}$ in experiments
represents evidence for a ``so-called back-stress, an elastic contribution to
the strain-hardening response.''
They did not explain this conclusion, but cite other observations that are
frequently cited as evidence for a back-stress.
For example, polymers exhibit a memory effect, returning to nearly their
undeformed shape when heated above $T_g$.
In Ref. \cite{hoy08} we showed that simulations reproduce this
effect.\cite{foot1}
They also allow the magnitude of the stress driving this recovery to be
evaluated directly.
We found that the stress is entropic in origin and only of order the
melt plateau modulus, rather than the much larger $G_r$.

Govaert et al. also argue that the Bauschinger effect
in oriented polymer glasses implies a back stress.
There is an asymmetry in the stress needed to deform the system, with a larger
stress required in the direction that would lead to an increase in orientation.
This phenomenon is also consistent with our simulations and physical picture.
When an oriented system is deformed in a way that reduces orientation,
the constraints of chain connectivity can be satisfied in many ways.
There is no need for the correlated rearrangements that occurred during 
orientation and the stress will be smaller.

\section*{\sffamily \large CONCLUSIONS}
\label{sec:conclusions}

Simulations were used to examine the relation between the flow stress and 
hardening modulus as temperature or strain rate was varied.
Contrary to statements in Ref. \cite{govaert08},
the results are completely consistent
with experiments.\cite{govaert08}
Results for both simulation and experiments show a linear relation between $G_r$
and $\sigma_{flow}$.
In both cases the slope $C_1$ and intercept $C_0$ depend on
whether $T$ or $\dot{\epsilon}$ is varied and $C_0$ is more positive
for varying temperature than varying rate.
The observations also seem to be more easily understood in terms of a strain
hardening model based on plastic deformation than an entropic model.

A qualitative explanation of the trends with temperature and rate was given
based on the previously observed correlation between increases in stress and
increases in the rate of plastic rearrangement.\cite{hoy06,hoy07,hoy08}
Since the size and complexity of plastic rearrangements increase with
strain, thermal activation may be more effective for the initial
flow stress than $G_r$.
This would lead the flow stress to vanish at a lower temperature than
$G_r$, implying a positive $C_0$ for variations with $T$.
In contrast, the value of $C_0$ would be negative for rate-dependent data
if the simpler activations involved in $\sigma_{flow}$ lead to a
stronger rate dependence.
The simulations are consistent with these predictions for $C_0$ and
most polymers have $C_0 <0$ for rate-dependent data.
Larger values of $C_0$ are found in experimental results for temperature
dependence,
although in some cases they appear to be negative.
This may reflect the contribution of strain softening to the measured
values of $\sigma_{flow}$.

Many aspects of strain hardening result from the orientation of
molecules that is produced by strain.
Any mechanism that depends on this orientation will have memory effects
and may exhibit a Bauschinger effect, but different models give
very different predictions for the magnitude of these effects.
In conventional entropic network models, orientation represents a
loss of entropy that leads to a direct entropic stress.
This entropic term is much smaller than the hardening modulus, but
comparable to the stress driving the shape memory effect.\cite{hoy08}
Our simulations suggest that the influence of orientation on $G_r$ is indirect.
As the orientation increases, larger non-affine displacements and more
breaking of van der Waals bonds are required to maintain chain connectivity.
The growing constraints of chain connectivity on rearrangements are related
to the decreasing entropy. However the magnitude of the stress is 
much higher than the plateau modulus because the energy dissipated by
rearrangements is set by the flow stress rather than $k_B T$.

\subsection*{\sffamily \normalsize ACKNOWLEDGMENTS}

We thank Michael Wendlandt and Leon Govaert for useful comments.
The simulations in this paper were carried out using the LAMMPS molecular 
 dynamics software (http://lammps.sandia.gov).  
This material is based upon work supported by the National Science
Foundation under Grant No. DMR-0454947 and through the MRSEC
Program under Award No. DMR05-20415.


\begin{thebibliography}{25}
\bibitem[{foo({\natexlab{b}})}]{foot1}
\bibinfo{note}{Govaert et al. stated \cite{govaert08} that the strain in our
  simulations relaxes completely when stress is removed, but this is not the
  case. As in experiments, the recovery is only a few percent until the sample
  is heated to nearly $T_g$.}
\end{thebibliography}
\end{document}